\documentclass[preprint,review,number,sort&compress,times]{elsarticle} 
\usepackage[T1]{fontenc}
\usepackage{graphicx,subfigure,xcolor}
\usepackage{here}
\usepackage{fixltx2e}
\usepackage[squaren,Gray]{SIunits}
\usepackage{amsmath,amssymb}
\usepackage{sistyle,wrapfig,float,array,comment,multicol}
\usepackage{epstopdf,booktabs}
\newcommand*\Laplace{\mathop{}\!\mathbin\bigtriangleup}
\usepackage{url,csquotes,caption,listings}
\usepackage[hidelinks]{hyperref}

\newcommand{\bs}{ \boldsymbol}
\newcommand{\rs}{ \mathrm}

\def \be{\begin{equation}}
\def \ee{\end{equation}}
\def \bse{\begin{subequations}}
\def \ese{\end{subequations}}

\begin{document}

\begin{frontmatter}

\title{Anode-metal drop formation and detachment mechanisms in liquid metal batteries}
\author[1,2]{Sabrina Bénard\corref{cor1}}
\author[1]{Norbert Weber}
\author[1]{Gerrit Maik Horstmann}
\author[1]{Steffen Landgraf}
\author[1]{Tom Weier}

\cortext[cor1]{Corresponding author. Helmholtz-Zentrum Dresden –
  Rossendorf, Bautzner Landstr. 400, 01328 Dresden,
  Germany. \textit{E-mail address}: sabrina.benard@ens-paris-saclay.fr (S. Bénard)} 

\address[1]{Helmholtz-Zentrum Dresden – Rossendorf, Bautzner Landstr. 400, 01328 Dresden, Germany.}
\address[2]{Ecole Normale Supérieure Paris-Saclay, 4 avenue des Sciences, 91190, Gif-sur-Yvette, France.}

\begin{abstract}
  We study numerically localised short circuits in Li$||$Bi liquid
  metal batteries. In the prototype of a classical, three liquid-layer
  system, we assume a perceptible local deformation of the Li-salt
  interface. We find that there always exists a critical current at
  which a Li-droplet is cut off from this hump, and transferred to the
  Bi-phase. In a second case, we assume that the molten Li is
  contained in a metal foam, and that a small Li-droplet emerges below
  this foam due to insufficient wetting. This droplet is deformed by
  Lorentz forces, until eventually being pinched off. Here, the
  critical current is slightly lower than in the three-layer system,
  and both, a droplet transfer and complete short circuits are
  observed. Finally, we discuss the relevance of our simulations for
  experimentally observed short circuits and non-faradaic Li-transfer.
\end{abstract}

\end{frontmatter}


\section{Introduction}
Liquid metal batteries (LMBs) have been proposed as cheap stationary energy
storage about ten years ago \cite{Kim2013b}. Built by a stratification
of two molten metal electrodes, separated by a fused salt electrolyte,
such cells offer extreme current densities, self-healing properties
and a potentially unlimited life-time. The ohmic resistance of the
electrolyte layer represents typically the highest overpotential of
the cell
\cite{Agruss1962a,Vogel1969,Cairns1973,Kelley2018}. Therefore, the
electrolyte needs to be as thin as possible. Still, it should be thick
enough to avoid any short circuit between the metal electrodes. This
optimisation problem leads to electrolyte layers, which are typically
5-10\,mm thick. Paste electrolytes, where the molten salt is
stabilised by a ceramic powder are usually only 1-3\,mm thick
\cite{Agruss1962,Shimotake1969,Vogel1970,Cairns1969b,Weaver1962}, but
exhibit a 2-4 times lower conductivity \cite{Kelley2018}.

Apart from corrosion issues, internal short circuits are one of the
main reasons for cell failure, and have been reported e.g. for K-Hg
\cite{Agruss1962a,Agruss1962}, Mg-Sb \cite{Bradwell2011,Kim2017},
Li-Se \cite{Cairns1970a,Gay1972}, Ca-Bi \cite{Kim2013a}, and Li-BiPb
cells \cite{Kim2018}. As LMBs are mainly operated as closed cells, it
has not always been easy to determine the reason for cell failure by
post-mortem analysis. Especially in Ca-cells, dentrite-like
intermetallic phases are known to grow into the electrolyte, finally
short-circuiting the cell \cite{Kim2013a}. In Li$||$Sn batteries,
Sn-rich particles have been observed inside the electrolyte
\cite{Yeo2018}. Finally, it has been reported for Li-based LMBs that
a low wettability of the current collector leads to droplet-shaped
liquid electrodes, which decrease the interelectrode distance thus
raising the risk for a short circuit. Especially when using a metal
foam as current collector -- which should absorb the molten Li -- a
low wettability will lead to free-moving Li droplets, which may
short-circuit the cell \cite{Gay1972}.
\begin{figure}[H]
\centering
\includegraphics[width=0.5\textwidth]{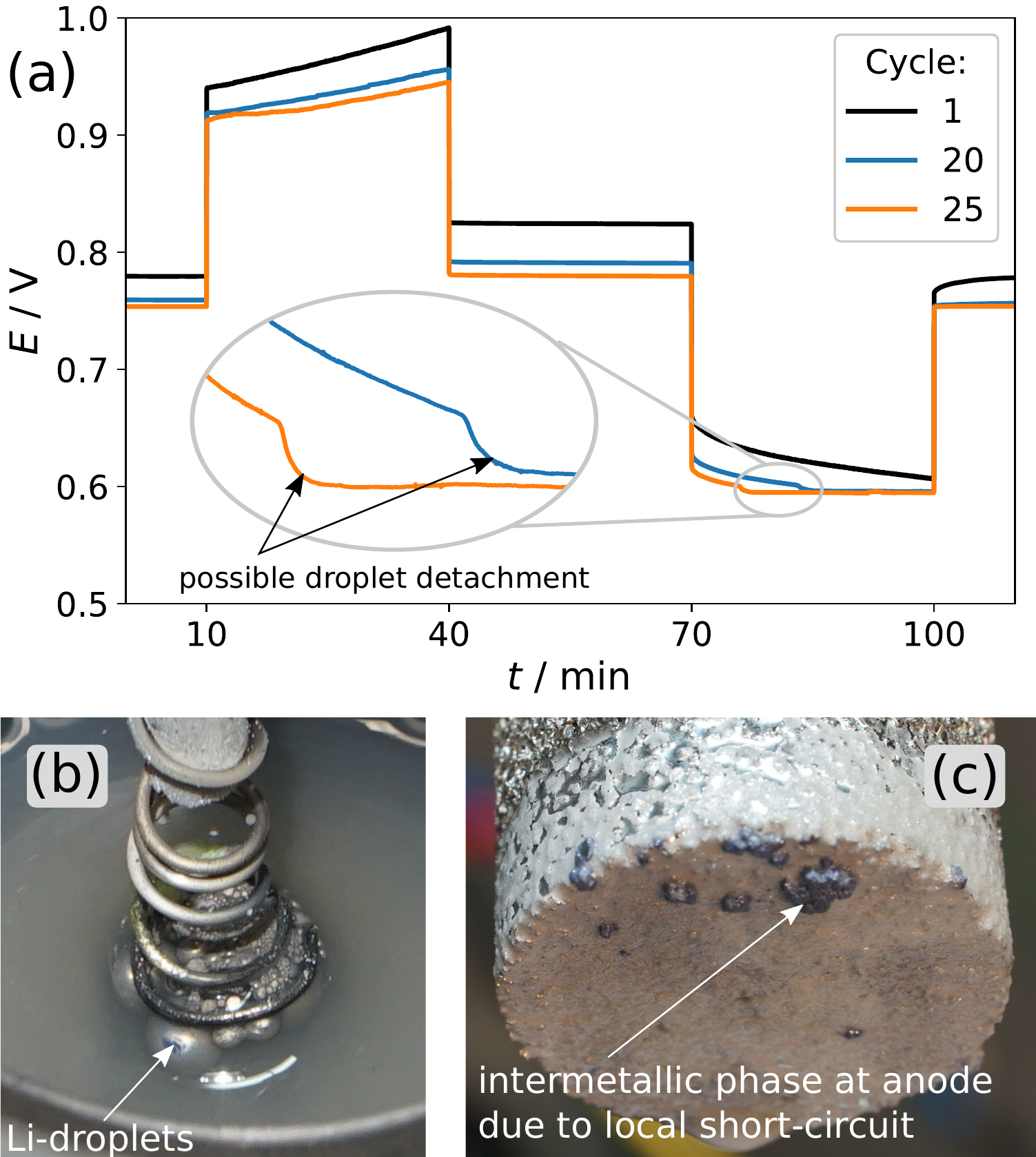}
\caption{Charge-discharge curve of a Li||Bi liquid metal battery with
  sudden voltage drops (a), Li-droplets at the negative current
  collector (b) and intermetallic Li$_3$Bi formed at the current
  collector after local short circuits (c). This suggests the existence 
  of droplet detachments and short circuits during operation.}\label{fig:1}
\end{figure}

Fig.~\ref{fig:1} illustrates localised short circuits observed in
Li$||$Bi cells. The inset in the charge-discharge graph shows a sudden
decrease in cell voltage -- an effect, which happens especially after
operating the cell for several cycles. We expect that such
voltage perturbations are caused by a short and non-faradaic transfer of
Li from the anode to the cathode. Thus, the Li-concentration in the
alloy will rise locally, therefore lowering the cell
potential. Similar effects with a typically larger signature in the
voltage profiles are known from vacuum arc-remelting
\cite{BeallBorgWood:1955} and electro-slag remelting
\cite{Campbell:1970}, where the droplet transfer from the melting
electrode to the pool is central to the refinement process.

One reason for the occurrence of sudden short circuits might be an
insufficient wetting of the current collector by liquid Li. This will
lead inevitably to the formation of Li-droplets during charge, as
illustrated in figure \ref{fig:1}b. If large enough, these droplets
might be transferred to the positive electrode by, e.g., the Lorentz
force or any fluid flow. A sufficient wetting is especially important
when using a metal foam to contain the molten Li. Unexpected surface
reactions, and missing wetting may lead to the formation of small
Li-droplets below of the foam. The following inevitable short circuit will not
only reduce the Coulombic efficiency, but might even lead to the
formation of solid intermetallic Li$_3$Bi. In the worst case, the
latter will attach to the metal foam -- as illustrated in
figure \ref{fig:1}c.

Inspired from aluminium-reduction cells, numerical modelling of liquid
metal batteries has focused in the past mostly on very large single
cells. Effects, which have been suspected to cause short circuits
include thermal convection
\cite{Shen2015,Koellner2017,Personnettaz2018a}, the Tayler instability
\cite{Stefani2011,Weber2013,Herreman2015}, electro-vortex flow
\cite{Weber2014b,Stefani2015,Ashour2017a,Herreman2019b,Liu2020} as
well as long-wave interface instabilities
\cite{Zikanov2015,Bojarevics2017,Weber2017a,Zikanov2017,Horstmann2017,Molokov2018,Tucs2018,Tucs2018a,Herreman2019,Xiang2019,Horstmann2019,Horstmann2020}.
All these flow phenomena will rather cause long-wave,
i.e. short circuits of the complete battery. They are probably not suited to
explain the experimental observations described before.

LMBs, which operate with three stratified liquid layers, i.e. which do not use a
metal foam to contain the anode metal, may suffer from localised
short circuits caused by short-wave interface instabilities. The
latter have already been studied in relation to electric-arc furnaces
by Sneyd \cite{Sneyd1985}. He found that local currents can be
destabilising if sufficiently strong. This work has later been
revisited by Priede \cite{Priede2016} in relation to LMBs. He found
that the electromagnetic force has a stabilising effect on short-wave
interface deformations, and that the latter are only unstable, if the
electrolyte has a higher conductivity than the electrodes. However,
this would be an atypical case. Of the more common electrode
materials, only Se has a sufficiently low conductivity. For very
strong, and non-linear interface deformations, also the kink- and
pinch-type instability of liquid columns might be of interest
\cite{IngardWiley:1962}.

The idea of a small Li-droplet, which forms due to insufficient
wetting below the metal-foam current collector resembles much the
mercury electrode used for electrochemical investigations
\cite{Newman2004}. As long as the droplet is still in contact with the
current collector, its surface will have a uniform potential. However,
as soon as the drop detaches, this is not longer the case. As surface
tension varies with the non-uniform potential, electrocapillarity
will possibly change the shape of the droplet and drive Marangoni
convection around it
\cite{Christiansen1903,Frumkin1945,Newman2004}. Electrocapillary
motion might therefore explain the movement of droplets, but not its
detachment.

A technical process, where droplet separation plays a crucial role, is
classical arc welding. There, the shape, size and detachment of the
fused metal droplets formed at the tip of the electrode depends
especially on the working current, i.e. the magnetic pinch force, and
the welding gas, which determines surface tension
\cite{Allum1985,Rhee1992,Nemchinsky1996}. The -- very important --
transition from globular transfer (of large droplets) to spray transfer (of
small droplets) has originally been explained by the static force
balance theory (SFBT). The latter predicts a drop detachment if the
retaining forces (surface tension) are larger than the separating
forces (gravity, Lorentz force)
\cite{Greene1960,Amson1965,Kim1993}. Alternatively, the pinch
instability theory (PIT) is used to explain droplet detachment
\cite{Kim1993}. There, a fluid column is expected to disintegrate into
a droplet chain due to the destabilising effect of the welding current
\cite{Allum1985}. The observation of pinch effects on current bearing
liquid columns dates back to Hering \cite{Hering:1907} and Northrup
\cite{Northrup1907}. Alpher \cite{Alpher:1960} gives an overview of
the older contributions.

Droplet formation at and detachment from electrodes is vital to two
metallurgical processes, namely vacuum arc remelting (VAR) and
electroslag remelting (ESR). Both processes are used for the
purification of metals that are melted by the passage of a large
current. Drops of molten metal detach from the electrode and pass
either through a vacuum (VAR) or through a slag layer (ESR) before
entering a melt pool that eventually undergoes solidification. Typical
current densities are 50\,kA\,m$^{-2}$ to 100\,kA\,m$^{-2}$ for ESR
\cite{KharichaWuLudwigRamprechtHolzgruber:2012} and between
1.5\,kA\,m$^{-2}$ to 600\,kA\,m$^{-2}$ for VAR \cite{Zanner:1979}. In both
processes the detaching drops leave a very distinct voltage signature
that is known as ``voltage or resistance swing'' in ESR
\cite{Kharicha2016}. Drop formation depends on
the shape of the electrode (from completely flat to very spiky) which
in turn is determined by the heat transfer conditions at the electrode
\cite{Korousic:1976}. Since the total current is usually high,
vigorous electro-vortex flows are to be expected in these
devices. They are strongest in the slag in case of ESR but --
naturally -- limited to the ingot for VAR \cite{Campbell:1970}. In
both processes droplets of high density material form at a downward
facing interface in a low density environment, i.e.~gravity can
reasonably be expected to be the main cause of drop detachment and
surface tension acts as a restoring force.

This is in sharp contrast
to the situation in liquid metal batteries considered here: the
droplets consisting of anode material possess a density lower than
that of their environment. Thus, surface tension as well as gravity
tend to keep the droplets at the anode. Obviously, the force balance
needs to be extended in order to provide an explanation for
droplet detachment. For this purpose we will focus in the following on
electromagnetic forces that were sometimes neglected
\cite{Campbell:1970,Korousic:1976} but recently gained more attention
\cite{KharichaWuLudwigRamprechtHolzgruber:2012,Kharicha2016}
in the frame of ESR models as well.

\section{Model and numerics}
We will present in this chapter the general model used for the
studies, including geometry, material properties and the governing
equations of the system. It will be applied in the next section to a 
Li|LiCl-LiF(70:30)|Bi liquid metal battery.

\subsection{Geometry}
\begin{figure}[!h]
\centering
\includegraphics[width=\textwidth]{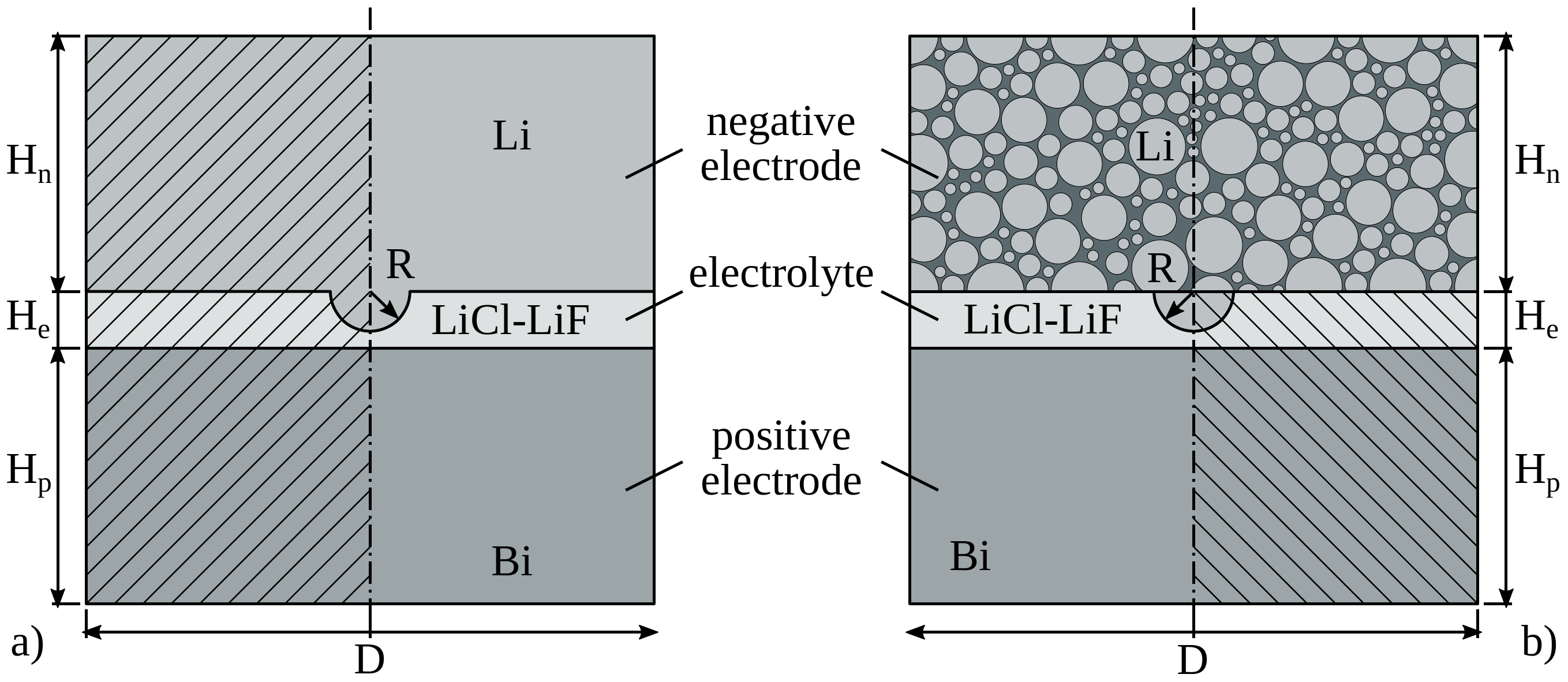}
\caption{Sketch of the three-layer cell (a) and the cell featuring a
  foam containing the positive electrode material (b). The interface
  deformation/droplet is located at the axis of symmetry.
  Numerical calculations of the fluid flow are limited to the hatched regions.}\label{fig:scheme}
\label{fig:2}
\end{figure}

The prototype used for the simulations is a cylinder. The diameter of
this cylinder is $D=10 \ \mathrm{cm}$, and the height of the three
layers are $H_p=4.5 \ \mathrm{cm}$, $H_e=1\  \mathrm{cm}$ and $H_n=4.5
\ \mathrm{cm}$ for the positive electrode, the electrolyte and the
negative electrode, respectively (figure
\ref{fig:scheme}). Generally, the negative electrode can consist of a
liquid metal layer, floating on top of the electrolyte, or the molten
Li may be contained in a metal foam. In order to study different
droplet-transfer mechanism, we will include both set-ups in our model,
denoting them as ``three-layer'' and ``foam'' case. In order to conduct our study, 
a hemispherical deformation will be taken as a perturbation in the first case,
and in the second case it will be a hemispherical droplet fixed below the foam. The
radius of the perturbation will be denoted by $R$.

The current collectors located above and below the battery will not be
modelled, but will be replaced by simplified boundary conditions. The
model will be set-up as being fully axisymmetric. 

\subsection{Material properties}
We assume a typical working temperature of 550 \degree C for the
Li|LiCl-LiF|Bi cell \cite{Ning2015}. The following table gives an
overview on the material properties.
\begin{multicols}{1}
\begin{table}[H]
\centering 
\begin{tabular}{llrrrrrrr}
\toprule
\multicolumn{1}{c}{Quantity} & \multicolumn{1}{c}{Unit} & \multicolumn{1}{c}{Li} & \multicolumn{1}{c}{salt} & \multicolumn{1}{c}{Bi}\\
\midrule
viscosity $\nu$    & $\mathrm{m^2\,s^{-1}}$& $6.3\cdot 10^{-7}$                & $1.6\cdot 10^{-6}$ & 1.2$\cdot 10^{-7}$\\
density $\rho$   &$\mathrm{kg\,m^{-3}}$& 482 &         
1568 &9721\\
electrical conductivity $\sigma$ &$\mathrm{S\,m^{-1}}$& 2.5$\cdot 10^6$ & 409 & 6.9$\cdot 10^5$\\
\midrule
  reference & & 
                \cite{Zinkle1998} &
                                    \cite{Janz1979} &
                                                      \cite{Sobolev2010,Alchagirov2016}\\
\bottomrule
\end{tabular}\end{table}
\end{multicols}
The interface tension $\gamma_{ij}$ between phases $i$ and $j$ is not
known; we use therefore the following values, which have been
estimated by Personnettaz et al. \cite{Personnettaz2018a} for a similar system:
\bse
\be
\gamma_{\rm \ Li,salt} = 0.196\, \mathrm{N\,m^{-1}},
\ee
\be 
\gamma_{\rm \ Bi,salt} = 0.275\, \mathrm{N\,m^{-1}},
\ee
\be
\gamma_{\rm \ Li,Bi} = 0.329\, \mathrm{N\,m^{-1}}.
\ee
\ese

\subsection{Equations}
While computing the Lorentz force, the magnetic field originating from
the feeding lines needs to be considered, as well. Simply neglecting
the field generated by currents outside the fluids would lead to a
reduced magnetic pressure towards the current collectors. The latter
would drive a nonphysical fluid flow. In order to avoid the costly
computation of additional magnetic fields, we employ the perturbed
current method \cite{Weber2017a}. We approximate the battery's
magnetic field as originating from an infinitely long cylinder,
denoted as $\bs{B_0}$. This allows to compute simply the perturbed
magnetic field $\bs{b}$. The total magnetic field in the battery will
thus be $\bs{B} = \bs{B_0} + \bs{b}$, and the current density is
written as $\bs{J} = \bs{J_0} + \bs{j}$ with $\bs{J_0}\ =\ -J_0
\bs{e_z} $. In order to find $\bs{B_0}$, we use Ampere's law 
\be 
\oint \bs {B_0} \cdot \bs {dl}\ = \ \iint \mu_0 \bs {J_0} \cdot  \bs {dS }\quad \Longleftrightarrow \quad B_0( r) \cdot 2\pi r\ =\ \mu_0 J_0\pi r^2,  \nonumber
\ee
with the vacuum permeability $\mu_0$. The magnetic field of the
infinitely long cylinder reads then 
\be 
\bs{B_0}(r) = \dfrac{\mu_0 J_0}{2} r \bs{e_\theta}.
\ee
The electric potential $\varphi$ in the battery is obtained by solving
the Laplace equation 
\be
\nabla \cdot (\sigma \nabla \varphi)\ =\ 0,  
\label{eq:potential}
\ee
which is derived by applying the divergence operator to Ohm's law 
\be
\bs J = \sigma \bs {E} = \sigma(-\nabla \varphi + \bs u \times \bs B),
\ee
while neglecting the term $\bs u \times \bs B$. The latter is
permissible, as the flow velocities are rather small. We find the
current density as 
\be
\bs{J}\ =\ -\sigma \nabla \varphi.
\label{eq:j}
\ee
The velocity $\bs u$ is obtained by solving the Navier-Stokes equations
\begin{subequations}
\be
\frac{\partial (\rho \bs{u})}{\partial t}\ +\ \nabla \cdot (\rho \bs{u} \bs{u})\ =\ -\nabla p_d\ + \ gz\nabla \rho\ + \ \nabla \cdot (\rho \nu (\nabla\bs{u}\ +\ (\nabla\bs{u})^\rs{T} ))\ + \ \bs{J}
	\times \bs{B} \ +\ \bs{f_{\rm st}},
\label{eq:ns}
\ee
\be 
\nabla \cdot \bs u\ =\ 0,
\ee
\end{subequations}
with $g$ denoting the gravity and $p_d= p-\rho gz$ the modified
pressure \cite{Rusche2002}. The surface tension force $\bs{f_{\rm st}}$ is
modelled using the continuum surface force  
(CSF) model \cite{Brackbill1992,Lam2009,Wang2015,Kissling2010},
i.e. it is implemented as
a volume force $\bs {f_{\rm st}} = \sum_i\sum_{j\neq
i}\gamma_{ij}\kappa_{ij}\bs {\delta_{ij}}$, concentrated at the
interface, with $\gamma_{ij}$ denoting the interface tension between
phases $i$ and $j$. The curvature is given as
\begin{equation}
\kappa_{ij}\approx-\nabla\cdot\frac{\alpha_j\nabla\alpha_i -
\alpha_i\nabla\alpha_j}{|\alpha_j\nabla\alpha_i - \alpha_i\nabla\alpha_j|},
\end{equation}
and the Dirac delta function $\bs{\delta_{ij}} = \alpha_j\nabla\alpha_i -
\alpha_i\nabla\alpha_j$
ensures that the force is applied only near an interface. Finally, we
find the surface tension force as
\begin{equation}
	\bs {f_{\rm st}} \approx -\sum_i\sum_{j\neq i}\gamma_{ij}
\nabla\cdot\left(\frac{\alpha_j\nabla\alpha_i -
\alpha_i\nabla\alpha_j}{|\alpha_j\nabla\alpha_i -
\alpha_i\nabla\alpha_j|}\right)(\alpha_j\nabla\alpha_i -
\alpha_i\nabla\alpha_j).
\end{equation}
Please refer to \ref{a:oscillation} for further details about
the surface tension and contact angle modelling. 

We find the induction equation for the magnetic field $\bs b$ by using
Faraday's and Ohm's law
\begin{subequations}
\be 
\nabla \times \bs {\delta e}\ =\ - \bs {\dot{b}},
\ee 
\be 
\bs j\ =\ \sigma \bs {\delta e},
\ee 
\end{subequations}
where $\bs {\delta e}$ denotes the perturbed electrical
field. Combining both equations leads to
\be 
\bs{\dot{b}}\ = \ -\nabla \times \bs {\delta e}\ = \ -\nabla \times \left( \frac{\bs j}{\sigma} \right) = \ -\nabla\left(\frac{1}{\sigma}\right) \times \bs{j} -\frac{1}{\sigma} \nabla \times \bs{j}.
\ee 
Applying Ampere's law as
\be 
\nabla \times \bs b\ =\ \mu_0 \bs j,
\ee 
we obtain
\be 
\nabla \times \bs j\ =\ \nabla \times \left( \frac{1}{\mu_0} \nabla \times \bs b \right) \ =\ \frac{1}{\mu_0} \left( \underbrace{\nabla(\nabla \cdot \bs b)}_{=\bs 0} - \Laplace{\bs b} \right) = - \frac{1}{\mu_0} \Laplace{\bs b}.
\ee 
Considering that the magnetic field varies slowly, i.e. $\bs{\dot{b}}
\approx \bs 0$, we can use the quasi-static approximation
\cite{Herreman2015,Weber2015b,Bandaru2016} and obtain the final
induction equation as
\be 
\frac{1}{\mu_0} \Laplace \bs{b}\ =\ \sigma \nabla\left(\frac{1}{\sigma}\right) \times \bs{j}.
\label{eq:b}
\ee
Due to the symmetry of the model, the only non-zero component of the
magnetic field is the azimuthal part $b_\theta$. Indeed, as the system
is axisymmetrical, every meridional plane containing the vectors
$\bs{e_z}$ and $\bs{e_r}$ is a symmetry plane for the current. Thus,
the magnetic field is necessarily perpendicular to this plane, that is 
to say only in the azimuthal direction.

Knowing current and magnetic field, the Lorentz force is computed as
\be 
\bs{f_{\rm L}}\ =\ \bs{j} \times \bs{B}\ +\ \bs{J_0} \times \bs{b}.
\ee 
Note that we do not include the constant part $\bs{J_0}\times\bs{B_0}$ as
it is rotation-free and can therefore be included into the pressure
gradient. As the magnetic field is purely azimuthal we find
\be
f_{\rm L,\theta} = J_z b_r - j_r b_z = 0.
\ee

Using a modified volume-of-fluid method
\cite{Hirt1981,Ubbink1997,Rusche2002}, the phase fractions $\alpha_i$
are solved as
\be 
\frac{\partial \alpha_i}{\partial t}\ +\ \nabla \cdot (\alpha_i \bs{u})\ =\ 0,
\ee 
using an approach similar to the Flux Corrected Transport method
\cite{Graveleau2016,Boris1973,Zalesak1979,Damian2013,Rudman1997}. The
mixture properties are then computed as
\be
\nu  =\frac{1}{\rho} \sum_i \alpha_i\rho_i\nu_i, \quad
\sigma = \sum_i \alpha_i\sigma_i, \quad 
\rho = \sum_i \alpha_i\rho_i.
\ee
We implement our model in the open source CFD library OpenFOAM
\cite{Weller1998}, using the existing solver
\textit{multiphaseInterFoam} \cite{Ubbink1997,Rusche2002}, where we
add equation \eqref{eq:potential}, \eqref{eq:j}, \eqref{eq:b}, as 
well as the Lorentz force in the equation \eqref{eq:ns} \cite{Weber2017a}.

\subsection{Boundary conditions}
For every simulation, we use non-slip boundary conditions for the
velocity as well as the corresponding Neumann boundary condition for
pressure. Biot-Savart's law might be applied to determine the correct
boundary values of the magnetic field as \cite{Meir1994,Weber2017b}
\be 
\bs b(\bs r) = \frac{\mu_0}{4\pi}\int \frac{\bs j(\bs   r')\times(\bs
  r - \bs r')}{|\bs r - \bs r'|^3}dV'.
\ee
However, as the borders are far from the deformation, the perturbation of
the magnetic field at the boundaries will be almost zero. We set
therefore $\bs{b} = \bs{0}$ on all boundaries, which
saves much time by avoiding the computation of Biot-Savart's law.

Finally, we fix the current density to $-J_0 \bs{e_z} $ on the top and on
the bottom of the battery, which is equivalent to fixing the gradient
of the potential, since we have $\nabla \varphi =\ - \dfrac{\bs{J}}{
  \sigma} $.

\subsection{Contact angle}
In order to model the Li-droplet in the foam case, we need to impose a
contact angle between the droplet and the foam. It has been shown that
the contact angle $\theta$ depends on several parameters such as the materials,
the surface roughness and the temperature \cite{Fiflis2014, Zuo2014,
  Krat2017}. Moreover, the foam is a porous surface which means that
the contact angle can take several values already because of its
porosity \cite{cassie1944}. Finally, the wetting properties of the
foam will also change during battery operation because of chemical
reactions between salt and foam. Consequently, we do not know the
exact value for the contact angle in our case. For this reason, we
will limit the investigation to a 90\degree\ contact angle that
approximates at the same time a mean value of the reported contact
angles.

The shape of the droplet can be obtained by solving the
Young-Laplace equation, which can be in our case re-written as 
\be
-\gamma \nabla \cdot \bs n = 2\gamma b + gz \Delta \rho,
\ee
where $\bs{n}$ is the unit normal vector pointing out of the surface
and $b$ the curvature at the bottom of the drop.

An algorithm provided by Korhonen \cite{Korhonen} allows us to compute
the analytical solution of the shape of a droplet with desired
material properties, contact angle and volume of the drop. We will use
this algorithm in order to validate the solver.

\section{Validation}\label{s:validation}
The magnetohydrodynamic model has already been validated by
comparison with an experiment \cite{Weber2017b}. Moreover, the
multiphase model has been compared successfully with analytical
results in the past \cite{Weber2017b,Herreman2019}. We will focus
therefore on the validation of the contact angle and on the droplet
shape, in the foam case.

\begin{figure}[H]
\centering
\includegraphics[width=0.5\textwidth]{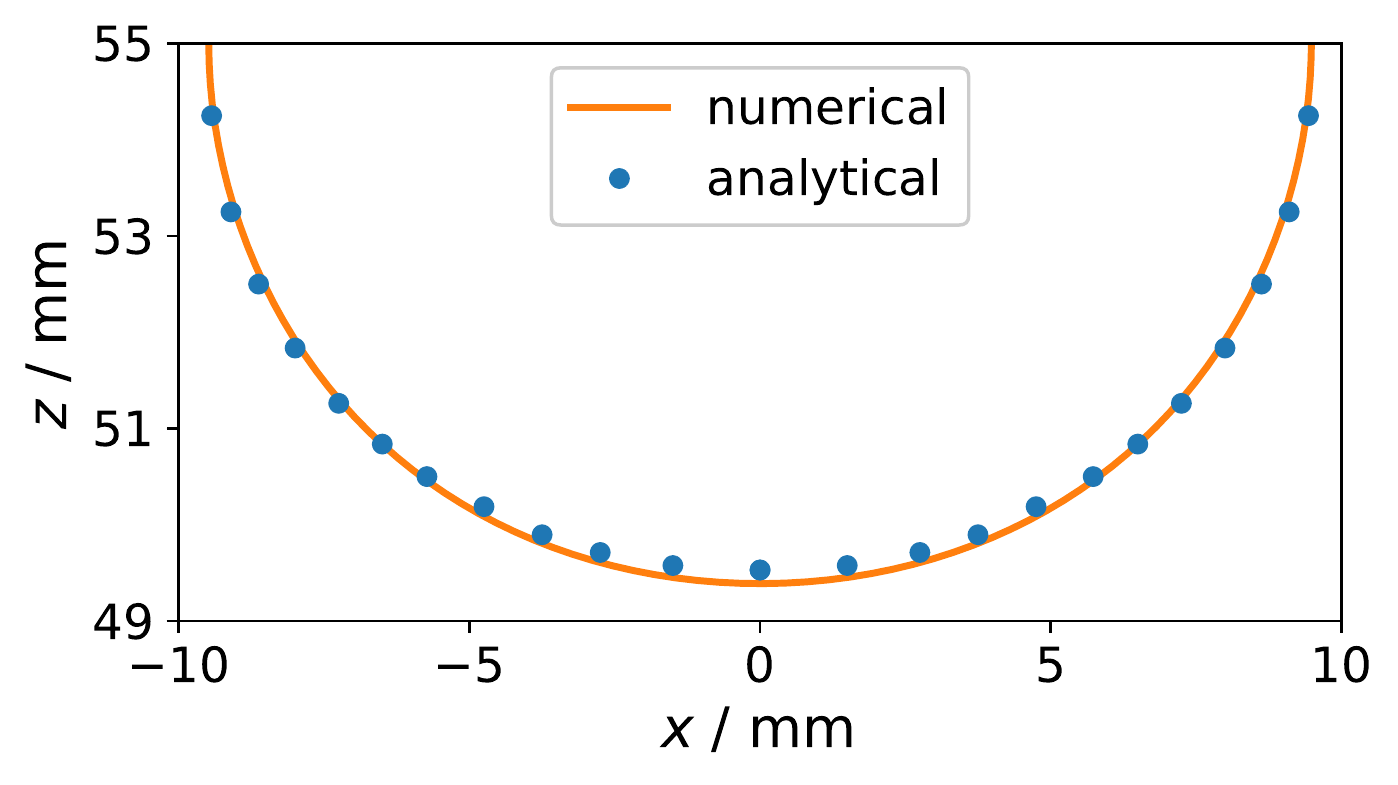}
\caption{Shape of a Li-droplet on a surface obtained by the analytical
  solution (blue) and the numerical solution (orange) for $\theta =
  90 \degree$ and $R = 8$ mm. Both shapes are quite well superposed, 
  which validates the solver.}
\label{fig:3}
\end{figure}

For this purpose, we model a Li-droplet with a volume corresponding to
the one of a half-sphere with a radius of 8\,mm and a 90\degree\
contact angle. After placing this initial drop under the metal-foam,
we run the simulation until the drop takes its final shape. Figure
\ref{fig:3} shows this contour, while comparing it with the analytical
solution obtained from Korhonen \cite{Korhonen}. Overall, both curves
match quite well. They deviate slightly only in the vicinity of the
contours turning point. We denote the error $\epsilon$ of the droplet
shape as \be \epsilon = \dfrac{h_{\rm a}-h_{\rm n}}{h_{\rm a}}, \ee
where $h_{\rm a}$ and $h_{\rm n}$ are the minimal height of the drop
of the analytical and numerical solutions, respectively. We find
$\epsilon=2.5\%$, which we consider acceptable in terms of accuracy of
our solver.

\section{Results and discussion}
\subsection{Three liquid layers}

As a first step, we study the three-layer case. A deformation of the
Li-salt interface is used as initial perturbation. The latter is
assumed to be of spherical shape in order to simplify the problem, and
is placed at the centre of the battery so that we can use
axisymmetrical simulations. Within the parameter studies, the radius of the deformation
and the current density are varied.

\begin{figure*}
\includegraphics[width=\textwidth]{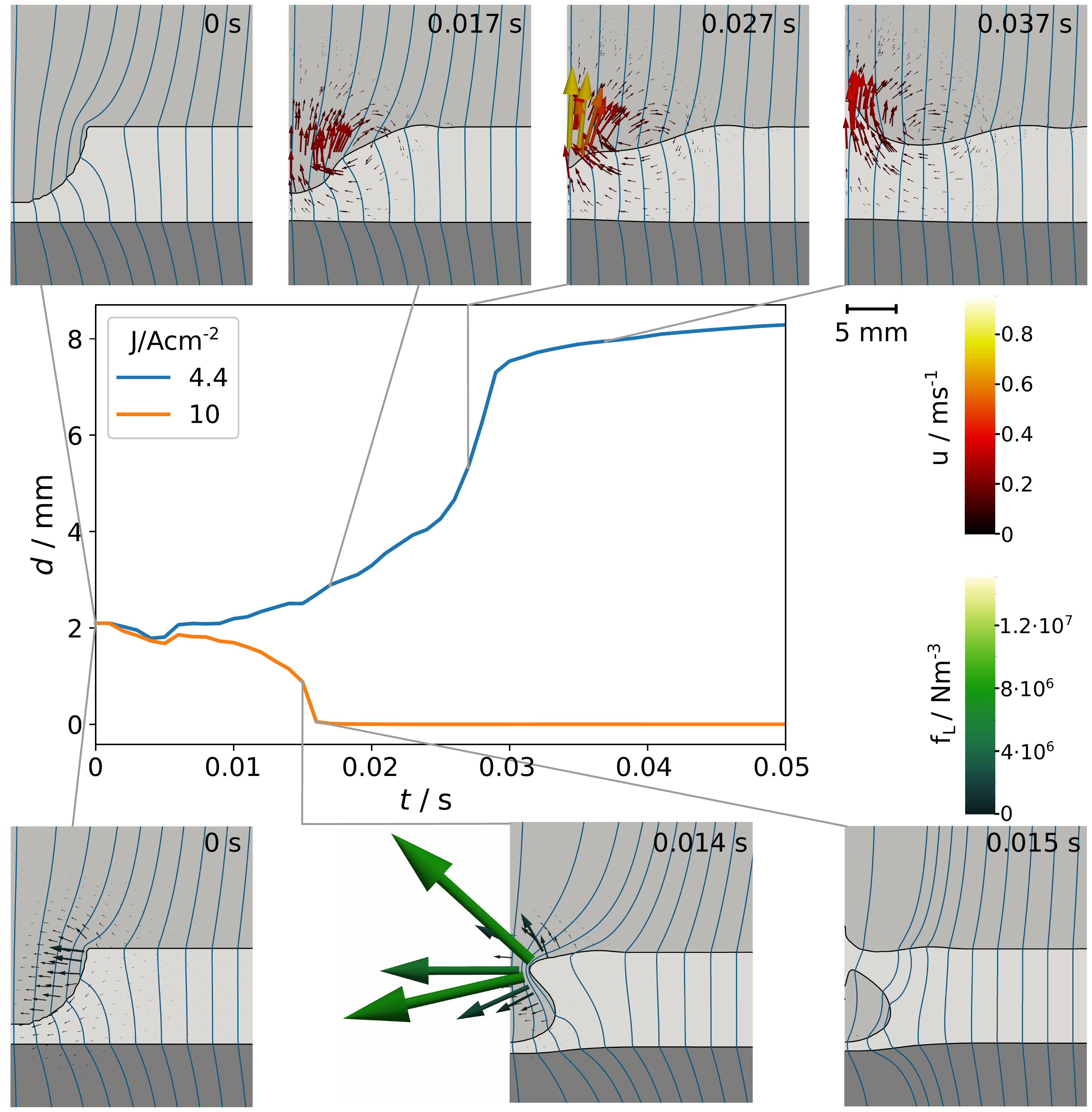}
\caption{Minimal
lithium-bismuth distance for the three-layer simulation vs. time
(central plot). The contour plots on the top show the phases (grey
values), current lines (blue) and velocity vectors ($u$) colored by
magnitude. Contour plots at the bottom display Lorentz force ($f_\mathrm{L}$)
instead of velocity. The insets show only a part of the cell around the deformation.
For a current density of 4\,A\,cm$^{-2}$, the system is re-stabilized by gravity. 
For a current density of 10\,A\,cm$^{-2}$, the system is destabilized by the Lorentz 
force with a Li-droplet detachment.}
\label{fig:4}
\end{figure*}

Figure \ref{fig:4} illustrates the time evolution of the interface for
a deformation with an initial radius of 8\,mm. In the upper insets, a
current density of 4.4\,$\mathrm{A\,cm^{-2}}$ has been applied. Here,
the system re-stabilises by gravity: all Li floats up, until the
deformation disappears. The same can be seen in the graph in the
middle of the figure, which shows the minimal distance between the Li and Bi phases
$d$ as a function of time. The blue curve, with a current density of
4.4 $\mathrm{A\,cm^{-2}}$, clearly confirms the re-stabilisation of the
system: the interelectrode distance increases with time until the
interface regains its equilibrium shape.

The lower insets illustrate a second case with a larger current
density of 10 $\mathrm{A\,cm^{-2}}$. Now, the resulting Lorentz force
is large enough to destabilise the system. The Lorentz force acts
predominantly at the upper right corner of the deformation as the
current lines are strongly bent in this area. It overcomes the
stabilising effect of surface tension, and shears off a Li-droplet
within only $\sim$0.015\,s. This very quick pinch creates an inward directed
flow, which might eventually push the droplet downwards to the
Bi-phase. This sequence of events is mirrored by the orange graph in
figure \ref{fig:4}, which shows how the Li-droplet approaches the Bi-interface
with time, until finally touching it. Please note that
we do not observe a complete short circuit, but just a droplet transfer
from the negative to the positive electrode. Secondly, our
axisymmetric model will probably not deliver a very realistic
description of the droplets downwards-movement, because of its
enforced axial-symmetry. The symmetry condition at
the axis deflects the inward directed flow down- and upwards. In
reality, this flow will probably be 3-dimensional. Nevertheless, we
expect our model to describe the initial pinch- and cut-off-phase very
well. 

The discussion of figure \ref{fig:4} clearly indicates the existence of
a critical current density, at which the system is destabilised, and a
droplet pinched off from the negative electrode. In order to obtain this
critical current density, a parameter study with different deformation radii
and current densities has been performed. The results are presented in table
\ref{tab:Cases}.  

\begin{table}[H]
\centering
\begin{tabular}{cc}
\toprule
\multicolumn{1}{c}{Droplet radius (mm)} & \multicolumn{1}{c}{Critical current density ($\mathrm{A\,cm^{-2}})$} \\
\midrule
3 & $14$ \\
4 & $11$\\
5 & $8.4$ \\
6 & $6.6$\\
7 & $5.5$\\
8 & $4.7$\\
9 & $4.1$\\
\bottomrule
\end{tabular}
\caption{Critical current density depending on the radius of the
  initial deformation. The electrolyte is always 10\,mm thick.}
\label{tab:Cases}
\end{table}

The table shows clearly that especially large deformations are prone
to droplet formation and transfer. However, even there the current
density is with 4\,A\,cm$^{-2}$ still four times higher than for typical
Li$||$Bi cells \cite{Cairns1967}. As quite violent flows will be
required to create such large initial perturbations, we conclude that
the appearance of short-wave instabilities is rather unlikely -- at
least for the Li$||$Bi system. This finding is perfectly in line with
Priede \cite{Priede2016}. Finally, we would
like to stress that even when a droplet cut-off appears, this does not
necessarily lead to a metal bridge between both
electrodes. Typically, the Li-transfer occurs only in the form of
droplets.\\[0.2cm]

\begin{figure*}
\centering
\subfigure[]{\includegraphics[width=0.5\textwidth]{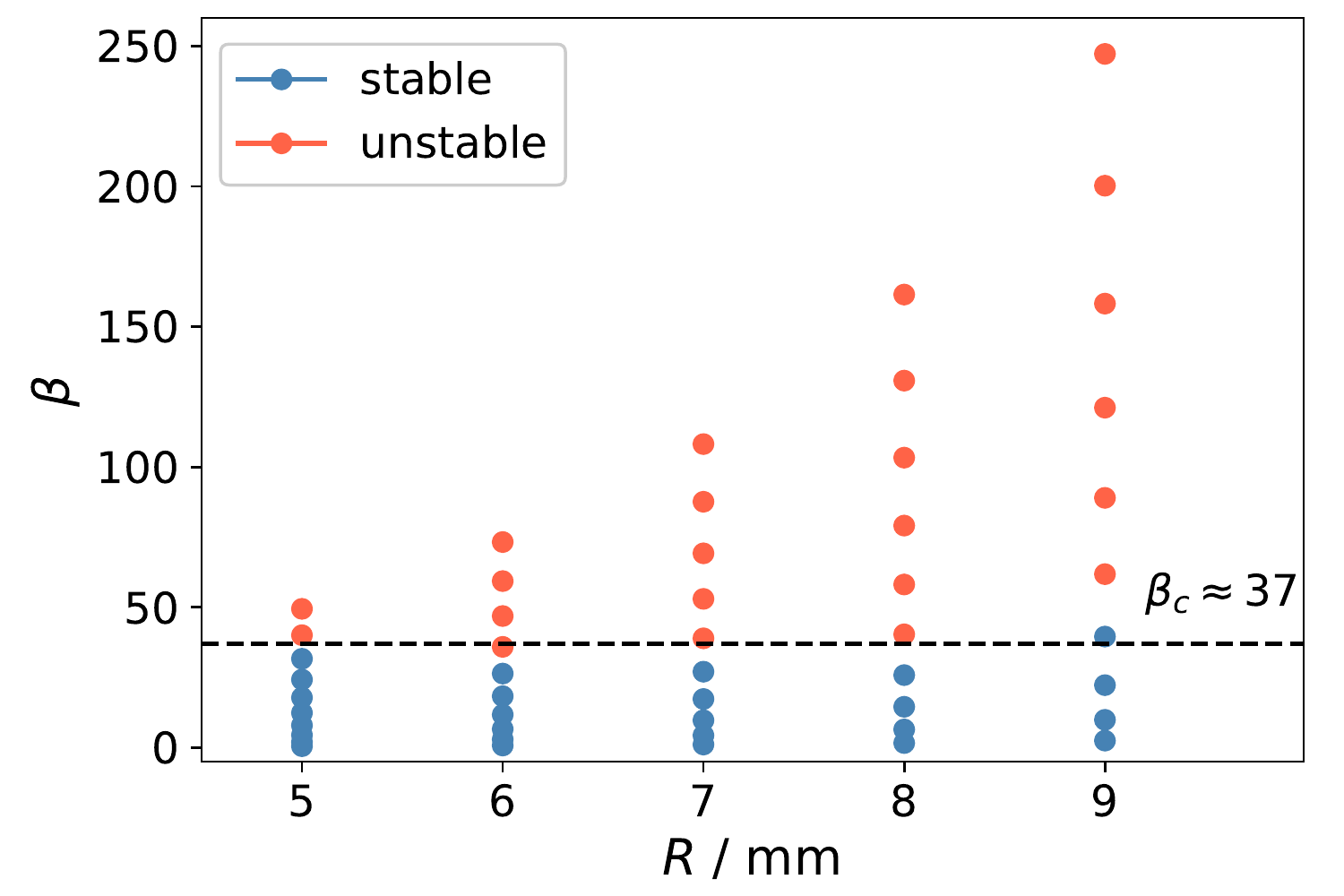}}
\subfigure[]{\includegraphics[width=0.5\textwidth]{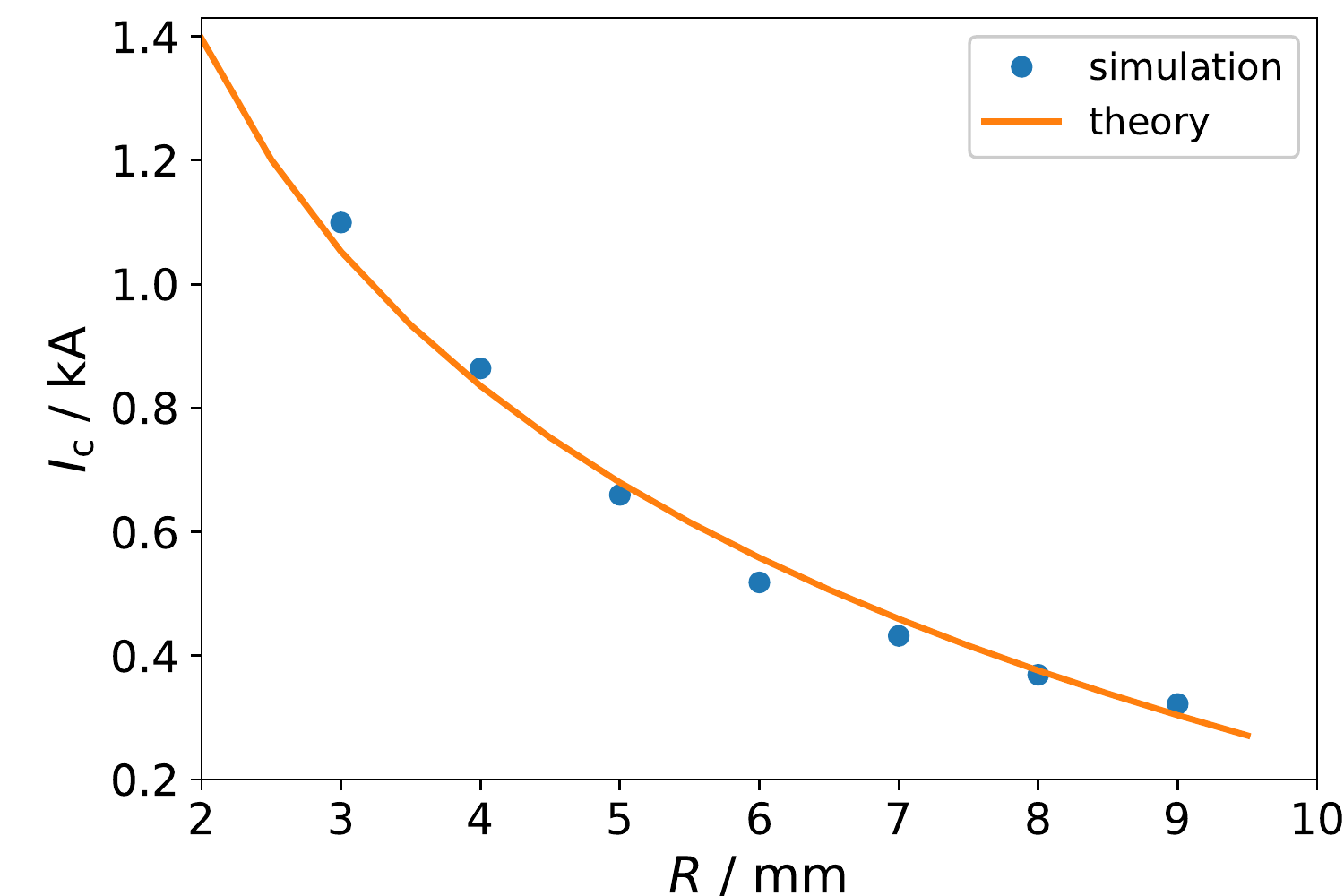}}
\caption{Stability diagram of the dimensionless parameter $\beta$ vs. $R$ (a) 
	and comparison between theoretical and simulated results of the 
	critical current vs. $R$ (b). This shows the existence of a 
	critical value of $\beta$ above which the system is destabilized, as well as a good
	agreement between theoretical predictions and numerical results.}
\label{fig:7}
\end{figure*}

In order to transfer our results to other battery chemistries and
geometries, it would be very useful to predict the critical cell
currents by a simple analytical formula. Therefore, we seek to develop a
first stability criterion in the following. The easiest way of
estimating the resilience of the interface against perturbations is to
compare the magnitude of the competing forces -- as it is done for arc
welding using the ``static force balance theory''
\cite{Greene1960,Amson1965,Kim1993}. In general, both gravity and
interfacial tension act as restoring forces, always counteracting
displacements with respect to the equilibrium positions, i.e.,
perfectly flat interfaces possibly involving menisci in the vicinity
of the contact lines. However, for predicting if spherical-like
deformations will detach or not, interfacial tension is by far higher
significance, because gravity, acting vertically, cannot compensate
the horizontally concentrated Lorentz force, see figure
\ref{fig:4}. Hence, we assume that only the interfacial tension force
$F_{\mathrm{st}}$, for spherical perturbations of radius $R$ simply scaling
as
\begin{equation}
	F_\mathrm{st} \sim \gamma_{\rm \ Li,salt}R,
\end{equation}
competes against the Lorentz force. The delicate task is now to model
the Lorentz force $F_\mathrm{L}$, which is induced by the interface deformation
itself. As the latter arises from the interaction of the cell current
with its self-induced magnetic field, it will be  of the order 
\begin{equation}
	F_\mathrm{L} \sim \mu_0 I_\mathrm{p}^2 , \label{eq:Lorentzforce}
\end{equation}
where $I_\mathrm{p}$ denotes the induced current contributing to the Lorentz
force. The cell current $I_0$, which is purely vertical in unperturbed
cells, only leads to a conservative Lorentz force distribution
balanced by the pressure. Any interface displacement will lead to a
redistribution of the cell current, which involves then a non-vertical
component $I_\mathrm{p}$, because the electrical resistance in the electrolyte
is orders of magnitude higher than in the liquid metals. Hence, we
need to calculate how strong $I_0$ is deflected by given perturbations
of the electrolyte -- similar as described in
\cite{Sneyd1985,Weier2020}.
Assuming that the lower interface remains flat and
$\sigma_\mathrm{n} , \sigma_\mathrm{p} \gg \sigma_\mathrm{e}$, the
horizontal perturbation current is of the order
\begin{equation}
I_\mathrm{p} \sim I_0 \frac{\eta}{H_\mathrm{e} -\eta}, \label{eq:CompCurrent}
\end{equation}
where $\eta$ is the characteristic depth of the perturbation. In the
considered case of spherical droplets, we can identify $\eta$ with the
mean height of the half sphere $\eta = 2R/3$. Inserting
(\ref{eq:CompCurrent}) in (\ref{eq:Lorentzforce}) finally yields the
scaling of the Lorentz force
\begin{equation}
F_\mathrm{L} \sim \frac{\mu_0 I_{0}^2 R^2}{(3H_\mathrm{e} -2R)^2}.
\end{equation}
As the last step, we can construct a dimensionless parameter by
comparing the forces $F_\mathrm{L}$ and $F_\mathrm{st}$ in the following
way:
\begin{equation}
	\beta := \frac{F_\mathrm{L}}{F_\mathrm{st}} = \frac{\mu_0 I_{0}^2}{\gamma_{{\rm \ Li,salt}}}\frac{R}{(3H_\mathrm{e} -2R)^2}. \label{eq:Criterion}
\end{equation}
Lithium drops are expected to detach from the anode layer if $\beta$ exceeds some critical threshold $\beta > \beta_c$. 
This statement is verified in figure \ref{fig:7} (a), showing the stability of the system when both the current and the radius of the simulated drop are varied. A sharp instability onset of
$\beta_c \approx 37$ becomes clearly apparent; only the largest $R =
9\, {\rm mm}$ sphere deviates significantly. The reason is most likely
that gravity cannot be neglected anymore for such large $R$ -- here we
reach the limits of the applied simplifications. However, all other
cases give evidence that $\beta$ predicts the interfacial stability
adequately. Knowing $\beta_c$, we can rearrange equation
(\ref{eq:Criterion}) to further predict the critical cell currents 
\begin{equation}
	I_\mathrm{c} = \sqrt{\frac{\beta_\mathrm{c} \gamma_{{\rm \ Li,salt}}}{\mu_0}}\frac{(3H_\mathrm{e} -2R)}{\sqrt{R}}, \label{eq:Ic}
\end{equation} 
above which the interface is unable to resist perturbations of size $R$. 
In figure \ref{fig:7} (b), we compare the simulated critical cell
currents against equation (\ref{eq:Ic}). Again, a good agreement with
respect to the uncertainties is evident, suggesting that the
simplistic stability parameter (\ref{eq:Criterion}) already reflects the
essential physics. Up to here, $\beta_c$ remains as an empirical
parameter. It is expected to highly depend on the geometrical cell
parameters, such as the cell diameter, and the material properties. A more rigorous analysis will
be necessary for an analytical prediction of $\beta_c$ -- a promising
task we leave open for future studies. 

\clearpage
\subsection{Metal foam electrode}
In this section we assume that the liquid Li, which forms the upper
electrode, is contained in a metal foam. We focus on the idea that a
small Li-droplet forms below of this foam due to insufficient wetting
between foam and Li. We investigate, if this droplet will be
transferred to the Bi-phase, or will keep its shape and position. The
size of the droplet will be varied as well as the current density,
exactly as in the previous section. 

All simulations are performed in two steps: first, the setup is
initialised with a hemispherical droplet of desired size and with a
contact angle of 90\degree. Then, we run the solver without Lorentz
force until the droplet takes its equilibrium shape. Finally, we
launch the solver another time, but now with the Lorentz force, in
order to observe a possible deformation or movement of the droplet.

\begin{figure*}
\centering
\includegraphics[width=0.5\textwidth]{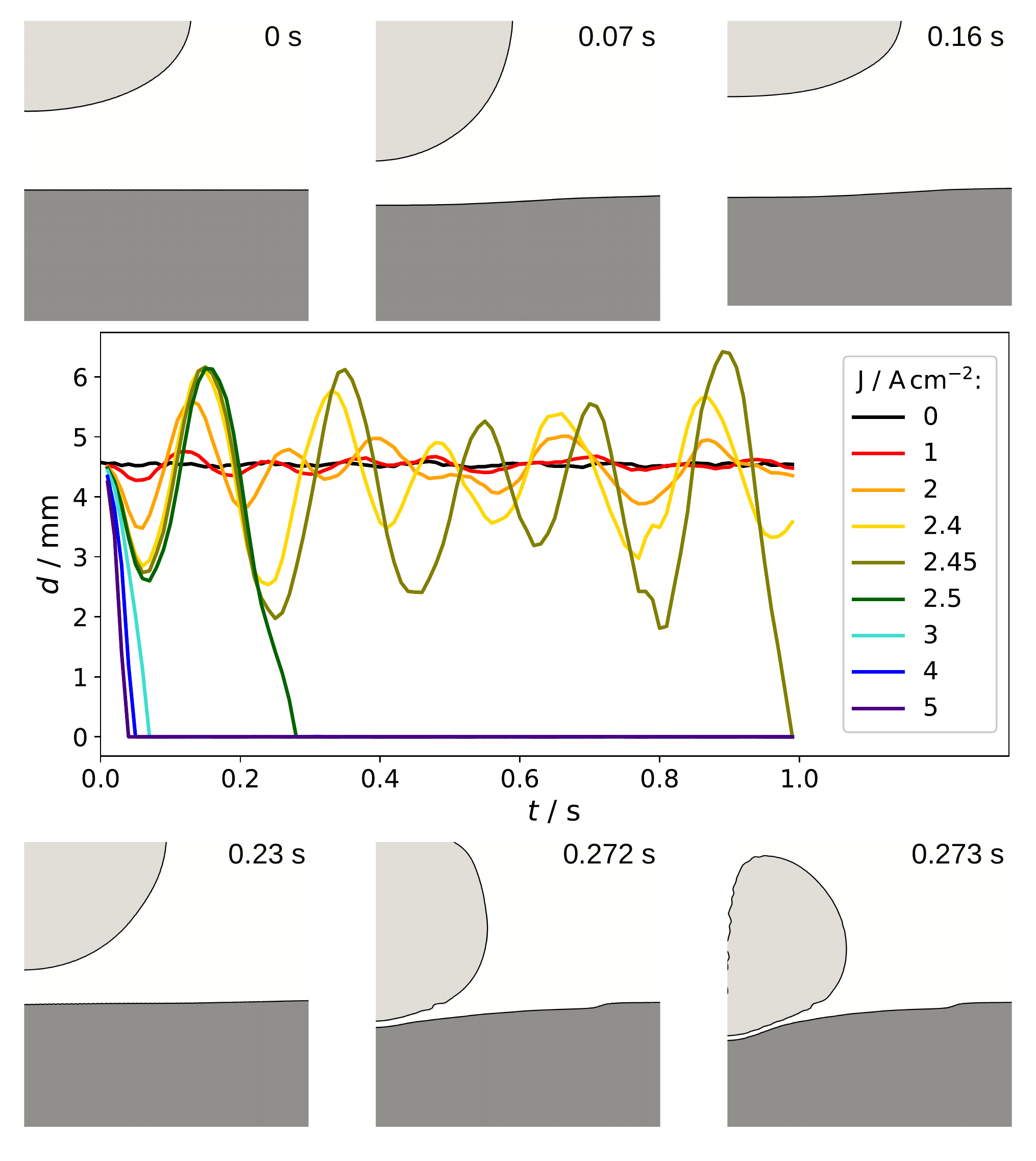}
\caption{Minimal distance between lithium and bismuth for different
  current densities and an initial droplet diameter of $R=8$\,mm. There
	exists a critical current densities above which the drop is detached 
	($d$ = 0). The insets illustrate the case $J=2.5$\,A\,cm$^{-2}$, 
	where the droplet oscillates before leading to a final
  short circuit and then detaching. The electrolyte layer is 10\,mm thick.}
\label{fig:5}
\end{figure*}

As a first case, we use a Li-drop with an initial radius of $R=8$\,mm.
Figure \ref{fig:5} illustrates the minimal distance between Li and
Bi over time. We see clearly that the drop oscillates, if the current
density is below of $J=2.45\,\mathrm{A\,cm^{-2}}$. While these
oscillations are dampened for low currents, they lead to some kind of
short circuit for current densities between 2.45 and 2.5
$\mathrm{A\,cm^{-2}}$. At even higher current, no oscillation appears,
and the interelectrode distance drops quickly to zero -- which
indicates an immediate short circuit, or a droplet cut-off and
transfer to the Bi-electrode.

The insets in the same figure illustrate the droplet shape over time
for $J=2.5\,\mathrm{A\,cm^{-2}}$, i.e. a case with initial
oscillation and final short circuit. When switching the current on,
the strong Lorentz force pinches the drop thus stretching it
downwards. Surface tension and gravity, trying to counteract this
movement, push the droplet again against the foam. These opposing
forces, and the hysteresis between them, explain perfectly the
oscillation, which finally leads to a short circuit after only
0.272\,s. Please note that the fairly large drop leads rather to a
direct contact of the Li- and Bi-phase than to droplet detachment and
subsequent transport.

\begin{figure*}
\centering
\includegraphics[width=0.5\textwidth]{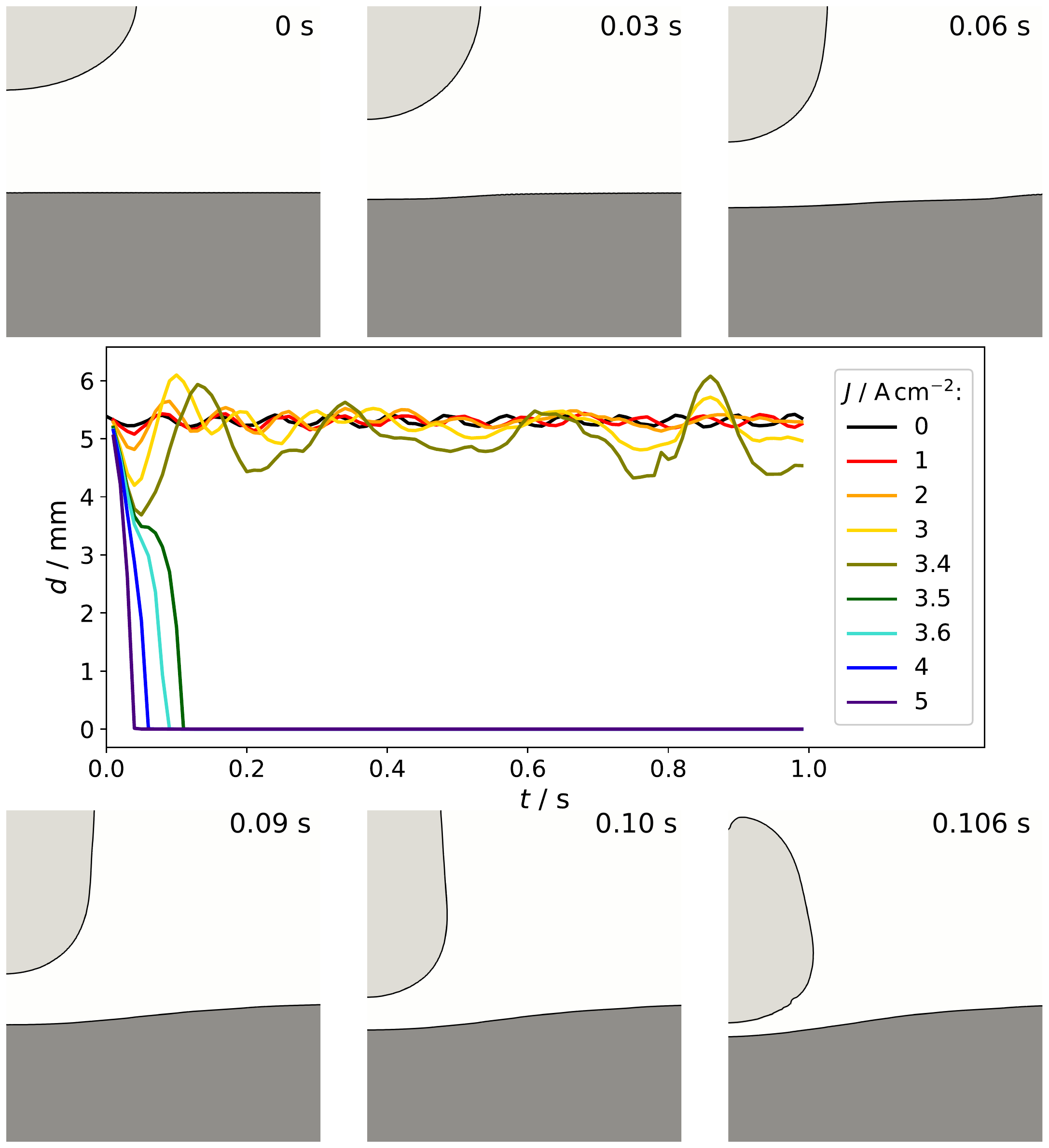}
\caption{Minimal distance between lithium and bismuth for different
  current densities and an initial droplet diameter of
	$R=6$\,mm. There
        exists a critical current density above which the drop is detached 
        ($d$ = 0). The insets illustrate the case $J=3.5$\,A\,cm$^{-2}$, 
        where the droplet oscillates before detaching but without short circuit. 
	The electrolyte layer is 10\,mm thick.}
\label{fig:6}
\end{figure*}

In a second case we study a smaller droplet with a radius of $R=6$\,mm, 
as illustrated in figure \ref{fig:6}. In this case, the critical
current density is $J=3.5\,\mathrm{A\,cm^{-2}}$, i.e. it is 40\%
higher than for the 8\,mm droplet. Yet here, the insets shown for
$J=3.5\,\mathrm{A\,cm^{-2}}$ indicate that neither an oscillation
nor a short circuit appear now. The droplet deforms, and simply
detaches, before being transferred to the Bi-phase.

Table \ref{tab:Icr:foam} summarises the different critical current
densities for radii in the range [6 mm, 9 mm]. Exactly as in the
three-layer case, smaller droplets need larger currents to
detach. However, the critical current densities are now lower than in the
three-layer case, and thus more realistic. Furthermore, we observe
short circuit already at a droplet radius of $R=7$\,mm -- which was
not the case in the three-layer simulations.
\begin{table}[H]
\centering
\begin{tabular}{cc}
\toprule
\multicolumn{1}{c}{Droplet radius (mm)} & \multicolumn{1}{c}{Critical current density ($\mathrm{A\,cm^{-2}})$} \\
\midrule
6 & $3.5$\\
7 & $3$\\
8 & $2.5$\\
9 & $2.3$\\
\bottomrule
\end{tabular}
\caption{Critical current density for droplet detachment for the foam case.}
\label{tab:Icr:foam}
\end{table}

\begin{figure*}
\centering
\subfigure[]{\includegraphics[width=0.5\textwidth]{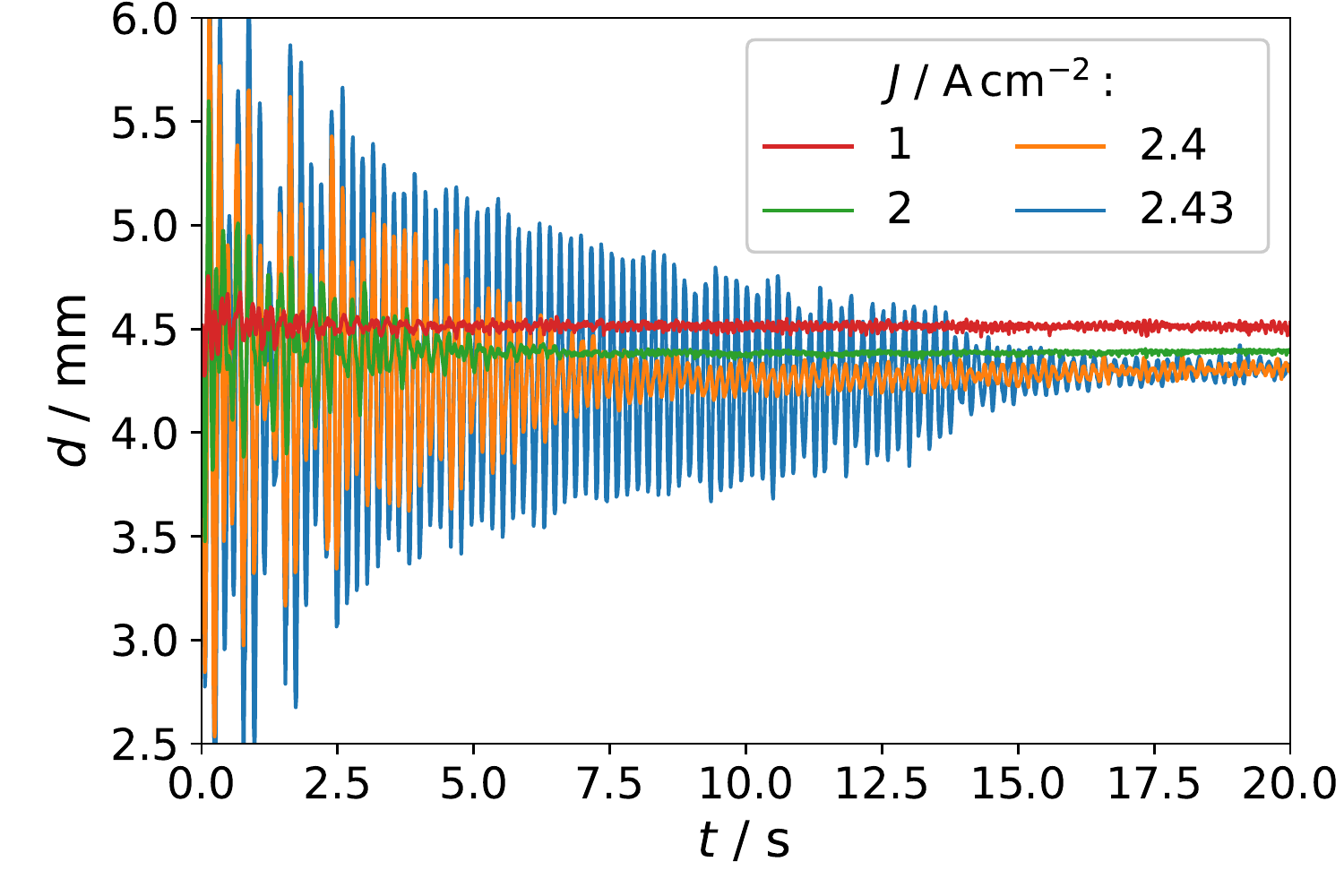}}
\subfigure[]{\includegraphics[width=0.5\textwidth]{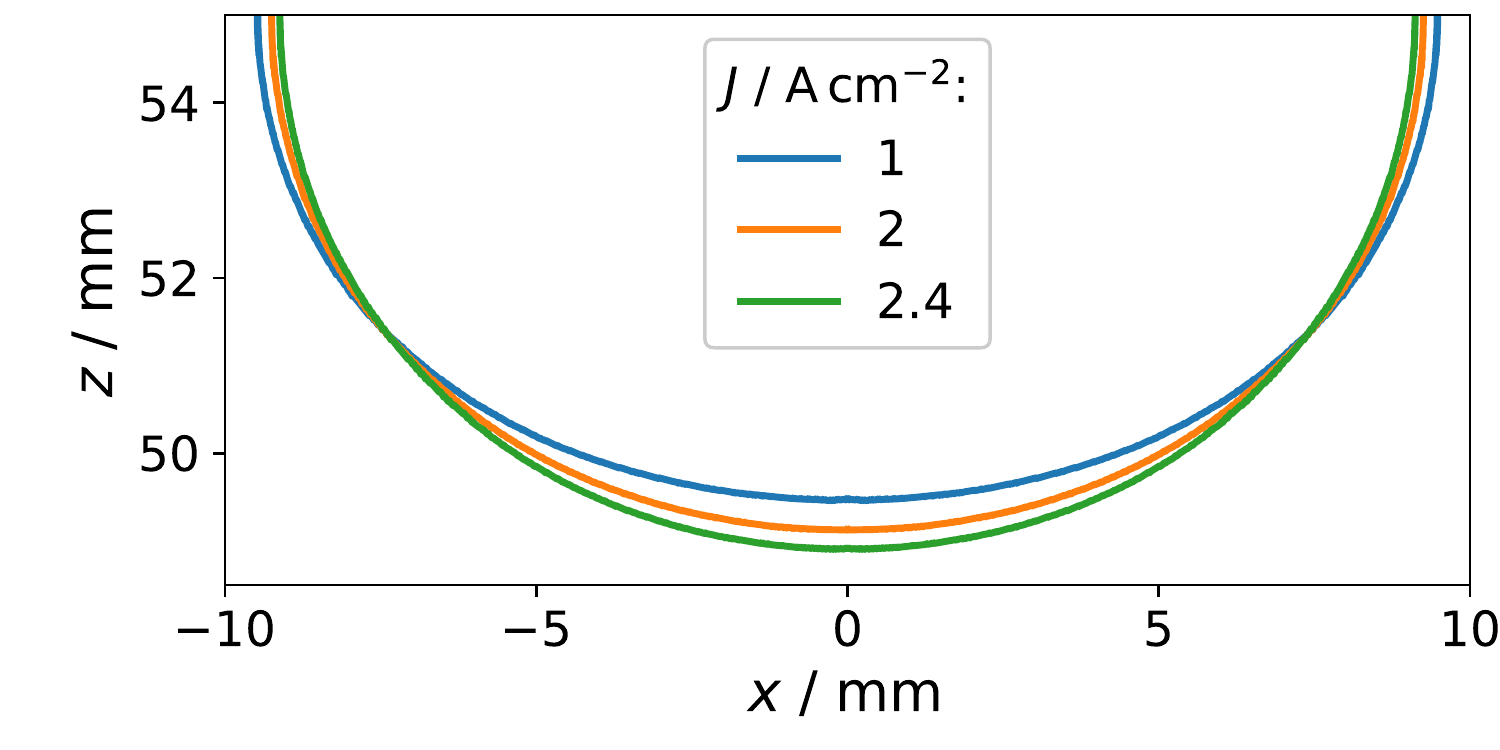}}
\caption{Minimal distance between lithium and bismuth (a), and
stationary droplet shape (b) for an initial droplet radius 
$R$ = 8\,mm. The electrolyte layer is 10\,mm thick. The amplitude of the
	oscillations decreases with time until it disappears.
	The final shape of the droplet depends on the current density.}
	\label{fig:8}
\end{figure*}

In order to investigate the nature of the droplet oscillation further,
we run the simulation for $R=8$\,mm until reaching a steady
state. Figure \ref{fig:8}a illustrates the time evolution of the minimal
interelectrode distance $d$ for different current densities. As
expected, the oscillation amplitude increases with current density,
and becomes significantly high when approaching the critical
current. In all cases, the oscillations are dampened with time until a
steady state is reached. Interestingly, the final interelectrode
distance depends on the current density, which means that the steady
shape of the droplet depends on the cell current. These ``final''
shapes are represented for three current densities in figure
\ref{fig:8}b. We can clearly see that larger current pinches the
droplet thus stretching it downwards. Apparently, the Lorentz force is
not strong enough to detach the drop, but sufficiently strong to
deform it -- leading to a new balance between the
forces.

In the foam case, we can observe short circuits, contrary to the three-layer case. The main difference between both cases is that in the three-layer LMB, the gravity force acts more significantly because the interface eletrolyte-negative electrode can move freely. In both cases, the droplet is stretched but in the three layer LMB, the system is initially unstable, the droplet being a deformation of the interface. Thus, the gravity force tries to make the droplet disappear because the instability of the system. In the foam case, the system is on the contrary initially stable, without Lorentz force. It means that the gravity force is larger in the three-layer case, and that is likely why in this case the droplet is cut before it touches the positive electrode. In the foam case, the gravity force acts, too, but not so significantly which allows in certain cases the droplet to short-circuit the cell before detaching.

Although we have shown that Lorentz forces can deform and even cut off
Li-droplets, it is not easy to generalise our findings. The problem is
simply too complex. While we used a 10\,mm-thick electrolyte, it might
be much thinner in other cases. While we assumed a contact angle
between droplet and metal foam of 90\degree, the latter might approach
180\degree\ in certain cases -- see figure \ref{fig:1} for an
example. Impurities, which attach to the interfaces, may alter the
interface tension considerably. If the metal foam is not filled with
Li any more, the current needs to follow the metal mesh. This might
lead to a very concentrated current entry into the droplet, and
therefore even stronger Lorentz forces. Finally, the electrolyte layer
will change its thickness during operation, and the droplet its
volume. All these effects, or uncertainties, make it difficult to
predict, when a droplet transfer from the negative to the positive
electrode will happen. These uncertainties notwithstanding, the
scenario sketched above plausibly explains the experimentally observed
effects illustrated in figure \ref{fig:1}.

\section{Summary}
Focusing on Li$||$Bi liquid metal batteries, we have
investigated localised short circuits, as well the transfer of
Li-droplets from the negative to the positive electrode. After giving
experimental evidence for such effects, we have studied two
paradigmatic cases.

Firstly, we focused on a classical, three-layer LMB, where we imposed
an initial deformation of the Li-salt interface. We find that there
exists always a critical cell current, for which the resulting Lorentz
force cuts off a Li-droplet. This drop is eventually transferred to the
Bi-electrode. As the necessary current is fairly high, and the initial
perturbation large, we conclude that such short-wave instabilities are
rather unlikely to occur in reality. Moreover, we have shown that
they do not lead to a localised short circuit, but only to a droplet
detachment and transfer.

Secondly, we considered an LMB, where the Li is contained in a current 
collector made of metal foam. Assuming that due to insufficient
wetting small Li-droplets form below of this foam, we studied the
influence of the cell current on this droplet. Here again, we found a
critical current, which leads to droplet detachment caused by the
pinching Lorentz force. At cell currents slightly below of the
threshold, we observed a droplet-oscillation leading finally to a
stable, stretched shape of the Li-drop. If the electrolyte is
sufficiently thin, even a direct metal bridge from the negative to
the positive electrode may appear. We believe that such effects offer
a plausible explanation for the experimental observations described in
the introduction.

An extension of this work would be the study of a larger range of
contact angles in order to understand its impact on the
destabilisation, and of thicknesses of the electrolyte in order to
find a minimal height of the electrolyte from which the
destabilisation does not occur anymore. 

\section{Acknowledgement}
This project has received funding from the European Union’s Horizon
2020 research and innovation programme under grant agreement No 963599 and was supported by a predoctoral scholarship of ENS Paris-Saclay, by
the Deutsche Forschungsgemeinschaft (DFG, German  Research
Foundation)  by  award  number  338560565 and in frame of the
Helmholtz - RSF Joint Research Group ``Magnetohydrodynamic
instabilities: Crucial relevance for large scale liquid metal
batteries and the sun-climate connection'', contract No. HRSF-0044 and
RSF-18-41-06201. Fruitful discussions with W. Herreman and C. Nore are
gratefully acknowledged. 

\appendix

\section{Nonphysical droplet oscillations}\label{a:oscillation}
When validating the model, and simulating a single droplet, we often
observed strong oscillations of the interface shape -- even if we did
not apply any Lorentz force. Such oscillations are clearly nonphysical,
and have been observed by others, before
\cite{Guilizzoni2011,Guilizzoni2012}. The reason for their appearance
might be related to the poor interface curvature calculation of the
CSF model, together with interFoam's contact angle model
\cite{Kunkelmann2011,Maurer2016,Horgue2014,Nieves-Remacha2015,Naoe2014}. 

The simplest remedy to eliminate the oscillation is increasing the
viscosity, as already reported by Guilizzoni
\cite{Guilizzoni2011,Guilizzoni2012}. Further, a relaxation of the
phase fraction or pressure, as well as a strong interface compression
will help to damp the oscillation. However, the interface compression
will deform the drop such that it fits less perfectly to its
analytical shape. 

It has turned out that the very best solution to suppress the
oscillation is computing the curvature by a smoothed phase-fraction
field \cite{Hoang2013}. In our simulations, two smoothing iterations
were sufficient.

\bibliographystyle{elsarticle-num}
\bibliography{references}
\end{document}